\documentclass[pra,twocolumn,showpacs,superscriptaddress,floatfix]{revtex4-1}
\usepackage{amsmath,amscd,amssymb,color}
\usepackage{graphicx,amsfonts,epsf}
\usepackage{epstopdf}
\usepackage[utf8]{inputenc}
\usepackage{float}
\usepackage{color,hyperref}
\definecolor{darkblue}{rgb}{0.5,0.0,0.0}
\hypersetup{colorlinks,breaklinks,
linkcolor=darkblue,urlcolor=darkblue,
anchorcolor=darkblue,citecolor=
darkblue,pdfauthor=KavitaDorai, pdftitle=KavitaDoraiRW}
\usepackage{multirow}
\usepackage{enumerate}
\usepackage{array}
\usepackage{xcolor}

\DeclareUnicodeCharacter{2212}{-}
\begin{document}
\title{Experimental quantum state transfer 
of an arbitrary single-qubit state on
a cycle with four vertices using a coined quantum random walk}
\author{Gayatri Singh}
\email{ph20015@iisermohali.ac.in}
\affiliation{Department of Physical Sciences, Indian
Institute of Science Education \& 
Research Mohali, Sector 81 SAS Nagar, 
Manauli PO 140306 Punjab India.}
\author{Kavita Dorai}
\email{kavita@iisermohali.ac.in}
\affiliation{Department of Physical Sciences, Indian
Institute of Science Education \& 
Research Mohali, Sector 81 SAS Nagar, 
Manauli PO 140306 Punjab India.}
\author{Arvind}
\email{arvind@iisermohali.ac.in}
\affiliation{Department of Physical Sciences, Indian
Institute of Science Education \& 
Research Mohali, Sector 81 SAS Nagar, 
Manauli PO 140306 Punjab India.}
\affiliation{Vice Chancellor, Punjabi University Patiala,
147002, Punjab, India}
\begin{abstract}
We experimentally demonstrate the transfer of an unknown single-qubit state from
Alice to Bob via a two-step discrete-time quantum random walk on a cycle with
four vertices on a four-qubit nuclear magnetic resonance
quantum processor. The qubits with Alice and Bob are
used as coin qubits and the walk is carried out on in a two-qubit 
`Gaming Arena'.  In
this scheme, the required entangled state is generated naturally via conditional
shift operators during the quantum walk, instead of being prepared in advance.
We implement controlled operators at Bob's end, which
are controlled  by Alice's coin qubit and arena qubits, in order
to reconstruct Alice's randomly generated state at
Bob's end.
 To characterize the state transfer process, we
perform quantum process tomography by repeating the experiment for a set of
input states $\{ \vert 0\rangle, \vert 1\rangle, \vert +\rangle, \vert -\rangle
\}$.  Using an entanglement witness, we certify that the quantum walk generates
a genuine quadripartite entangled state of all four qubits.  
To evaluate the efficacy of the transfer scheme, We use quantum
state tomography to reconstruct the transferred state by calculating the
projection of the experimentally reconstructed four-qubit density matrix onto
three-qubit basis states.  Our results demonstrate that the quantum circuit is
able to  perform quantum state transfer via the two-step quantum random walk
with high fidelity.
\end{abstract} 
\maketitle 
\section{Introduction}
\label{sec1}
Quantum analogs of classical random walks are a versatile tool to perform
quantum information processing tasks such as universal quantum
computing~\cite{childs-prl-2009}, quantum search~\cite{shenvi-pra-2003} and
quantum simulation~\cite{rebentrost-njp-2009}, and have been comprehensively
reviewed in several
articles~\cite{ambainis-ijqi-2003,roysoc-2006,venegas-qip-2012}.  Quantum walks
have been experimentally implemented on several physical platforms such as
nuclear magnetic resonance~\cite{du-pra-2003,ryan-pra-2005,du-pra-2010}, trapped
ions~\cite{xue-prl-2009,ahringer-prl-2010}, atoms in an optical
lattice~\cite{genske-prl-2013} and
photonics~\cite{broome-prl-2010,alberto-science-2010}.  Quantum teleportation
refers to sending an unknown quantum state between two parties, Alice and Bob,
using classical information and quantum
correlations~\cite{bennett-prl-1993,briegel-prl-1998,gottesman-nature-1999}.
Teleportation was first experimentally demonstrated using a pair of entangled
photons~\cite{bouwmeester-nature-1997},  while later experiments used other
physical platforms such as trapped ions~\cite{barrett-nature-2004}, photonic
qubits~\cite{jin-natpho-2010}, liquid state nuclear magnetic resonance
(NMR)~\cite{nielsen-nature-1998}, superconducting qubits~\cite{bau-prl-2012},
and nitrogen vacancy centers~\cite{pfaff-science-2014}.

Standard quantum teleportation schemes require an entangled state to be shared
{\em a priori} between the two parties and and since in coined quantum random
walks the conditional shift operator can introduce entanglement between the coin
and position spaces, the question arose as to whether quantum random walks could
be used for quantum teleportation.  To resolve this question, a new scheme to
perform teleportation of an unknown quantum state was designed which used
two-step quantum walks on a line and on a cycle with four
vertices~\cite{wang-qip-2017}. Two quantum walkers on different quantum
structures such as a line, a cycle and two-vertices complete graphs with loops
were used to teleport an arbitrary two-qubit state and were generalized to $N$
walkers and $N$-qubit states~\cite{qinp-2019,qinp-2021,qinp-2022-secret}.  
Controlled quantum teleportation was 
proposed based on three-coin quantum random walks, where a conditional shift
operator was used to generate entanglement between the position and coin
spaces~\cite{qinp-2023-teleport}.

Experimental realizations of the quantum
teleportation of an arbitrary two-qubit state using a four-qubit cluster state
and of quantum teleportation using coined quantum walks were demonstrated on the
IBM cloud quantum computing 
platform~\cite{chatterjee-qip-2019,qinp-2020-panigrahi}.  An
arbitrary two-qubit quantum-controlled teleportation scheme wherein the sender
performs two Bell state measurements and the receiver performs a unitary
operation to reconstruct the state was performed on an IBM quantum
computer~\cite{chinesephys-2022}.  Perfect state transfer of entangled states by
quantum walks with two coins was demonstrated on the IBM quantum
experience~\cite{qtmsci-2019}.

In this work, we experimentally demonstrate the transfer of an unknown
single-qubit quantum state between two parties (Alice and Bob), via a quantum
random walk on a cycle with four vertices, on a four-qubit NMR quantum
processor.  Although previous quantum random walk schemes have been used to
achieve quantum teleportation, we choose to instead use the scheme to achieve
quantum state transfer. This is because our quantum processor of choice is based
on liquid-state NMR, wherein the qubits are physically located on the same
molecule, and it is hence natural to conceptualize the protocol as a state
transfer instead of state teleportation. We re-cast the quantum random walk
scheme as a two-player quantum game played by Alice and Bob in a two-qubit
`Gaming Arena', with Alice and Bob each having one coin qubit which they flip
and use to play the game. The standard protocol to transfer an unknown quantum
state via a two-step quantum random walk involves the generation of a shared
entangled state during the walk, projective measurements which are performed by
Alice and communicated to Bob via a classical channel, and finally a set of
measurements that Bob performs based on Alice's measurement outcomes.  Since
projective measurements are not possible on NMR quantum processors, while
controlled operations between all the involved qubits are possible, we use
unitary operations at Bob's end which are controlled by Alice's qubit and the
Arena qubits, to achieve the state transfer.  Our scheme  uses conditional shift
operators on the `Arena qubits' in the `Gaming Arena' to execute one step of the
quantum walk. After Alice and Bob execute one step each of the quantum random
walk by using their coin qubits, controlled operations are applied on Bob's coin
qubit, which  then gets pushed into the randomly generated initial state of
Alice's coin qubit.  This completes the quantum state transfer protocol.  We
note here that this scheme to transfer the state, which was randomly generated
by Alice's coin qubit, to Bob's end is implementable only because all the four
qubits (the two coin qubits as well as the two `Arena' qubits) are all located
within the same molecule.  We use quantum process tomography to characterize the
teleportation scheme, by repeating the experiment for a set of four different
input states  $\{\vert 0\rangle$, $\vert 1\rangle$, $\vert +\rangle$ and $\vert
-\rangle\}$.  The tomographed four-qubit state contains information about
four-qubit entanglement which is generated during the quantum walk.  We use an
entanglement witness to verify that the generated four-qubit state indeed has
genuine quadripartite entanglement~\cite{tokunaga-pra-2006}.  To verify that the
state has indeed been transferred with high fidelity, we reconstruct the state
by taking projections of the experimental four-qubit density matrix onto
three-qubit basis states which are determined from Alice's measurements.  We are
able to achieve near-complete transfer of the unknown quantum state with high
fidelity.  Our results also demonstrate the viability of our four-qubit NMR
system as a robust and versatile quantum processor, which has good decoherence
properties and a high degree of quantum control.

This paper is organized as follows: The basic theoretical framework is briefly
reviewed in Section~\ref{sec2}, with a review of the coined discrete time
quantum random walk and a description of how to accomplish transfer of an
unknown single-qubit state by quantum random walks on a cycle
with four vertices, contained
in Sections~\ref{sec2.1} and \ref{sec2.2}, respectively.  Details of the
experimental implementation of the teleportation protocol on an NMR quantum
processor are given in Section~\ref{sec3.1}.  Section~\ref{sec3.2} contains the
details of the experimental settings and NMR pulse sequences, while
Section~\ref{sec3.3} contains a detailed analysis of the experimental
results.
Section~\ref{sec4} contains a few concluding remarks.
\section{Basic Theoretical Framework}
\label{sec2}
In this section, the basics of modeling a discrete time quantum random walk on
four vertices using multiple coins as well as the protocol for 
transfer of
an unknown single-qubit state using a quantum random walk will be discussed.
\subsection{Coined Quantum Random Walks on a Closed Cycle}
\label{sec2.1}
A coined discrete time quantum random walk occurs in the joint space of position
(in which the `walker' walks) qubits and the coin (or coins) qubit(s).  
The total Hilbert space
can be written as $\mathcal{H}=\mathcal{H}_P\otimes \mathcal{H}_C$, where
$\mathcal{H}_P$ is the Hilbert space of the walker 
and $\mathcal{H}_C$ is the Hilbert space of the coin. 
At each step of the quantum walk, 
the coin is flipped, 
and then depending upon the state of the coin, the walker moves either in
the clockwise or the anticlockwise direction.  
If the coin is in the state
$\vert 0 \rangle$, the walker moves in the anticlockwise direction and if the
coin is in the state $\vert 1 \rangle$, the walker moves in the clockwise
direction. 
Since the quantum walk occurs over the closed topology  of a 
cycle, the walker moves in the clockwise (anticlockwise) direction and
not in the usual left(right) directions.
For a coined quantum walk on a closed cycle denoted by a graph $G(V,E)$ with
edges $E$ and vertices $E$, the Hilbert space $H_{v}$ is spanned by the vertex
states $\vert v \rangle$, where $v \in V$ and the coin Hilbert space $H_{c}$ is
spanned by the qubit states $\{ \vert 0 \rangle, \vert 1 \rangle\}$, with the
total Hilbert space being given by $H = H_{v} \otimes H_{c}$.  One step of the
quantum walk is controlled by the unitary operator $U = S(I \otimes C)$, where
$C$ is the coin flipping operator and $S$ is the conditional shift operator
acting on $H_{v} \otimes H_{c}$
which is defined as~\cite{wang-qip-2017}:
\begin{eqnarray}
S_{\pm} &=& \vert{i\pm 1\,\text{mod}\, n}\rangle \langle i\vert \nonumber \\
S &=& \sum_{i=0}^{n-1}(S_+ \otimes \vert 0 \rangle_c\,{}_{c}\langle 0 
\vert +S_-
\otimes \vert 1 \rangle_c\,{}_{c}\langle 1 \vert ) 
\label{eq1}
\end{eqnarray}
where $n$ refers to the number of vertices on the cycle.
We use a closed cycle with four vertices to implement a two-step
quantum random walk, with each vertex denoting a possible position. 
The position qubits change their location depending on the outcome of
the coin qubit after it is flipped.
Without loss of generality, the coin operator 
$C$ can be arbitrarily set to any single-qubit
unitary, without affecting the outcome of the quantum 
walk~\cite{wang-qip-2017}.
\subsection{Transferring a Single-Qubit State  on a Four-Cycle}
\label{sec2.2}
\begin{figure}[h]
\includegraphics[scale=1]{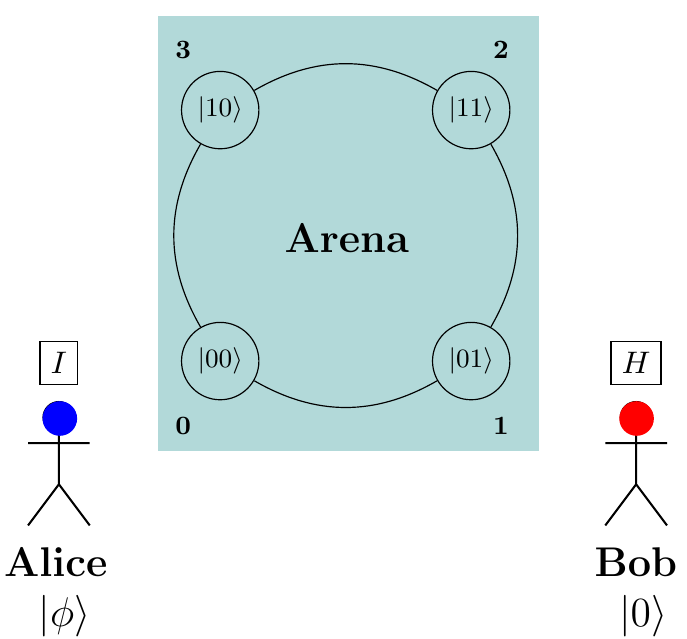}
\caption{The two-player `Game Arena' 
is depicted by a closed cycle with four vertices (${\bf
0,1,2,3}$) representing the states $\vert 00\rangle,\vert 01\rangle,\vert
11\rangle$ and $\vert 10\rangle$, respectively.  Alice and Bob use their
respective coin qubits to play the game and move from one vertex to another on
the closed cycle.  If the coin qubit is in the $\vert 0\rangle$ state, the move
on the closed cycle is executed in the anticlockwise direction, and if it is in
the $\vert 1\rangle$ state, the move is executed in the clockwise direction.
Alice's coin operator is the Identity $I$ gate, while Bob's coin operator
is the Hadamard $H$ gate. Alice begins the game with her coin in the
randomly generated $\vert \phi \rangle$ state, while Bob begins the game with
his coin in the $\vert 0 \rangle$ state.
}
\label{fig1}
\end{figure}

Standard teleportation schemes begin with preparing an entangled state to be
shared beforehand between the two parties. On the other hand, when using a
quantum random walk to achieve teleportation, the entangled state is generated
by two conditional shift operators during the steps of the walk.  
Previous NMR experimental protocols to realize quantum teleportation were
able to teleport the state of a nuclear spin ``locally'' from one atom to
another, located physically on the same molecule.  Since in the NMR scenario,
the state is teleported to another location within the same molecule, it can be
construed as not strictly being teleportation but as being akin to state
transfer.  We hence recast the entire theoretical scheme described in the
Reference~\cite{wang-qip-2017} in terms of achieving quantum state transfer
instead of quantum teleportation.  Consider a quantum random walk on a closed
cycle with four vertices, involving two coins, one with Alice and the other with
Bob.  The protocol is designed to transfer the coin state of Alice to the coin
state of Bob, and is similar in spirit to the quantum teleportation procotol.
The protocol can be visualized as a two-player game that Alice and Bob play,
with the quantum walk taking place in the space of the two `Arena
Qubits'. The schematic of the two-player game is depicted in Figure~\ref{fig1}.
The `Arena' space is spanned by the vertices and the coin space is spanned
by the edges of the cycle. There are two directed edges at each vertex in
Figure~\ref{fig1}. Thus, the dimension of the `Arena' space is four, while that
of each of Alice's and Bob's coin space is two.  
Initially, Alice's coin is prepared in an unknown
random one-qubit state denoted by $\vert \phi\rangle=a\vert 0 \rangle+ b \vert
1\rangle$ with $\vert a\vert^2+\vert b\vert^2=1$ in Figure~\ref{fig1}.  This
amounts to preparing the entire system in the initial state $\vert\psi_{{\rm
in}}\rangle=\vert 00\rangle \otimes \vert\phi\rangle\otimes \vert 0\rangle$.
Alice then uses her coin qubit $A_c$, to move one step in the random walk by
using the shift operator (denoted by the box labeled $W_1$ in
Figure~\ref{fig1}).  Bob starts with his coin in a state which is prepared by
applying a Hadamard gate on his qubit in the $\vert 0\rangle$ state, which is
thus in an equal superposition of $\vert 0\rangle$ and $\vert 1 \rangle $
states.  Bob then executes the next step of the quantum random walk by using his
coin and applying the conditional shift operator to the `Arena qubits' in the
game arena (denoted by the box labeled $W_2$ in Figure~\ref{fig1}).  After that,
controlled operations are applied to Bob's coin qubit.  Bob's coin is thus
pushed into the unknown random initial state of Alice. This step completes the
implementation of the protocol.  
The conditional shift operator (Eq.~(\ref{eq1}))
becomes
\begin{eqnarray}
S &=& (\vert 0\rangle \langle 1\vert +\vert 1\rangle \langle 2\vert+\vert
2\rangle \langle 3\vert+\vert 3\rangle \langle 0\vert ) \otimes \vert 1
\rangle_c \,{}_c\langle 1 \vert \nonumber \\ && 
+ (\vert 0\rangle \langle 3\vert+\vert
1\rangle \langle 0\vert +\vert 2\rangle \langle 1\vert 
+\vert 3\rangle \langle
2\vert) \otimes \vert 0 \rangle_c\,{}_c\langle 0 
\vert 
\label{eq2}
\end{eqnarray}
Without loss of generality, 
the coin operator of Alice is chosen to be the Identity
operation and the coin operator of Bob 
is chosen to be the Hadamard gate.

The state after the second step
of the quantum walk evolves to:
\begin{equation}
\vert\psi\rangle_2=\frac{1}{\sqrt{2}}(a \vert 0001 \rangle+b\vert 1000 \rangle
+ a \vert 0110\rangle+b \vert 1111\rangle)
\end{equation}
Alice now perform measurements on her qubits: on the position qubit in
the $\{\vert 0 \rangle, \vert 1 \rangle\}$ basis and coin qubit in the $\{\vert
+ \rangle,\vert - \rangle\}$ basis. 

Alice's measurement outcomes and Bob's
corresponding controlled operations
are given in Table~\ref{table1}. 
After two steps of the quantum walk, 
a Hadamard gate is applied on Alice's coin space
so that Alice performs all measurements in the $\{ \vert 0
\rangle, \vert 1 \rangle\}$ basis. Thus, the state after two steps
of the quantum walk is
given by:
\begin{eqnarray}
\vert\psi\rangle_f&=&\frac{1}{2} \{\vert 000 \rangle\otimes(a \vert 1 \rangle+b
\vert 0\rangle)+ \vert 100 \rangle\otimes(a \vert 1 \rangle-b \vert 0\rangle)+
\nonumber \\
&& \vert 011 \rangle\otimes(a \vert 0 \rangle+b \vert 1\rangle)+ \vert 111
\rangle\otimes(a \vert 0 \rangle-b \vert 1\rangle\} \label{eq4}
\end{eqnarray}
From the state $\vert\psi\rangle_f$ (Eq.~(\ref{eq4})), one observes that, if
Alice's position qubit is in the $\vert 00 \rangle$ state and Alice's coin qubit
is in the $\vert 0 \rangle$ state, Bob is required to apply a $\sigma_x$  $(X)$
gate
in order to recover the transferred state, and so on. 
After implementing the 
controlled operations 
listed in Table~\ref{table1}, the
state is given by:
\begin{equation}
\vert\Psi\rangle=\frac{1}{2} \{\vert 000 \rangle+ \vert 100 \rangle+ \vert 011
\rangle+ \vert 111 \rangle\}\otimes(a \vert 0 \rangle+b \vert
1\rangle)\label{eq5}
\end{equation}
Finally, by measuring Bob's qubit, one can verify that the state
$\vert\phi\rangle=a \vert 1 \rangle+b \vert 0\rangle$ has indeed been
transferred from Alice to Bob.  

\begin{figure}[h]
\includegraphics[scale=1]{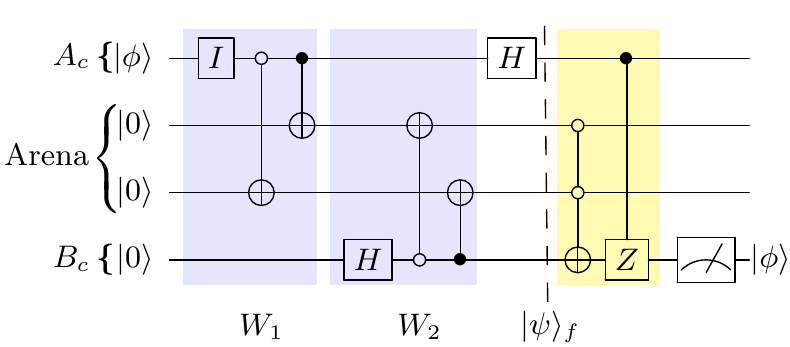}
\caption{Quantum circuit to transfer an
unknown single-qubit state $\vert\phi\rangle$
using a two-step quantum random walk 
on a cycle with four vertices. $A_{c}$ and $B_{c}$ denote the coin qubits
of Alice and Bob, respectively, while the two middle qubits are labeled
as the `Arena' qubits.
$\rm{W}_1$ and
$\rm{W}_2$ denote the first and second steps of 
the quantum walk. 
The yellow shaded box represents the implementation of unitaries
corresponding to the controlled
operations.
}
\label{fig2}
\end{figure}

The quantum circuit to experimentally realize the 
transfer of an unknown
single-qubit state using a two-step quantum random walk on a cycle with four
vertices is shown in Figure~\ref{fig2}.  
$A_c$ and $B_c$ denote the coin qubits of 
Alice and Bob, respectively, while
the other two qubits are denoted as the `Arena qubits'. The coin state
of Alice is the randomly generated single-qubit state $\vert \phi \rangle$, 
which is transferred to the coin state of Bob, at the end of the
circuit.
The
boxes labeled $W_1$ and $W_2$ contain the gates required to implement the first
and second steps of the quantum walk;   $I$ and $H$ and $Z$
denote an Identity operator, a Hadamard gate, and a Pauli $Z$ operator,
respectively.  The last yellow shaded box in Figure~\ref{fig2} contains the
gates required to implement the controlled
operations given in
Table~\ref{table1}. 

\begin{table}[h!]
\setlength{\tabcolsep}{12pt} 
\renewcommand{\arraystretch}{1.2}
\footnotesize{
\begin{tabular}{c c c}
\hline \hline
\multicolumn{2}{c}{Measurement Results} & Controlled Operation \\
$A_{c}$ & Arena Qubits &  $M$ \\
\hline \hline 
0&11& $I$\\
1&11& $Z$\\
0&00& $X$\\
1&00& $ZX$
\\
\hline \end{tabular} }
\caption{Measurement results on Alice's  coin qubit ($A_{c}$) 
and the corresponding controlled
operations $M$ on Bob's coin qubit ($B_{c}$); $Z$ and $X$ 
denote the Pauli matrices $\sigma_z$ and $\sigma_x$
respectively, while $I$ is the Identity (`do nothing') operator.}
\label{table1}
\end{table}
\section{Experimental implementation of teleportation via a
two-step quantum random walk} 
\label{sec3}
\subsection{Experimental NMR Details} 
\label{sec3.1}
\begin{figure}[t]
\centering
{\includegraphics[scale=1]{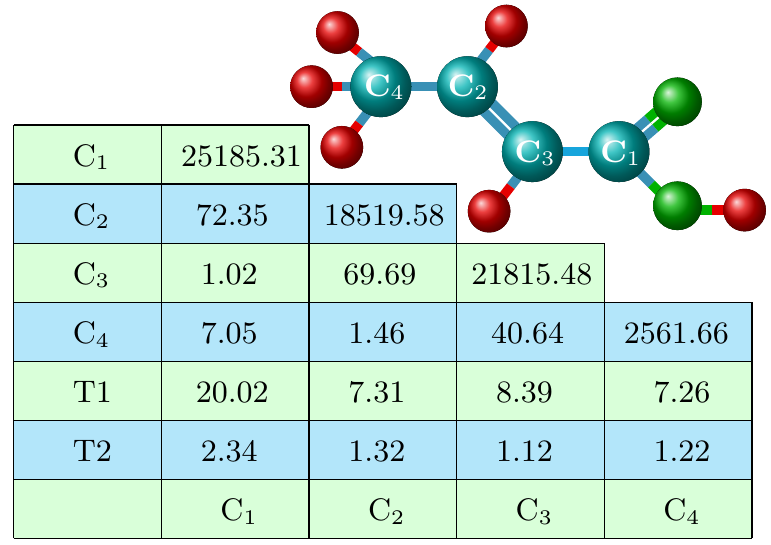}}
\caption{Molecular structure 
and system parameters of ${}^{13}$C-labeled trans-crotonic acid with
the four carbon atoms labeled as C$_1$,C$_2$,C$_3$,C$_4$.
The diagonal elements in the table represent the chemical shifts (in
Hz), while the off-diagonal elements are the J-couplings (in Hz)
between two spins. The T$_1$ and T$_2$ relaxation times
(in sec) are given in the last two rows of the table. 
Green and red balls represent oxygen and hydrogen atoms, respectively.}
\label{fig3}
\end{figure}

We used four ${}^{13}$C nuclei of the ${}^{13}$C-labeled trans-crotonic acid
molecule dissolved in acetone-D6 to physically realize four qubits.  The
molecular structure and system parameters including the chemical shifts $\nu_i$,
T$_1$ and T$_2$ relaxation times and the scalar J-coupling constants are given
in Figure~\ref{fig3}.  All the experiments were performed at ambient temperature
($\approx 300$K) on a Bruker DRX Avance III 600 MHz NMR spectrometer equipped
with a standard 5 mm QXI probe.  The methyl group and other proton spins were
decoupled throughout the experiments, via a composite pulse-based broadband
decoupling sequence (WALTZ-16). 

Under the weak coupling approximation, the NMR Hamiltonian for a four
spin-1/2 system in the rotating frame is given by:
\begin{equation}
\mathcal{H}=-\sum_{i=1}^{4} 
(\omega_{i}-\omega_{{\rm rf}}) I_{i z}+
\sum_{i<j,j=1}^{4} 2 \pi J_{i j} 
I_{i z} I_{j z}
\label{eq6}
\end{equation}
where $i$ labels the spin, $\omega_{{\rm rf}} $ is the rotating frame frequency,
$\omega_i=2\pi\nu_i$ and $I_{iz}$ denote the Larmor frequency and the
$z$-component of the spin angular momentum of the $i$th spin respectively, and
$J_{ij}$ is the strength of the scalar coupling between the $i$th and $j$th
spins.

\begin{figure*}[t]
\centering
{\includegraphics[scale=1]{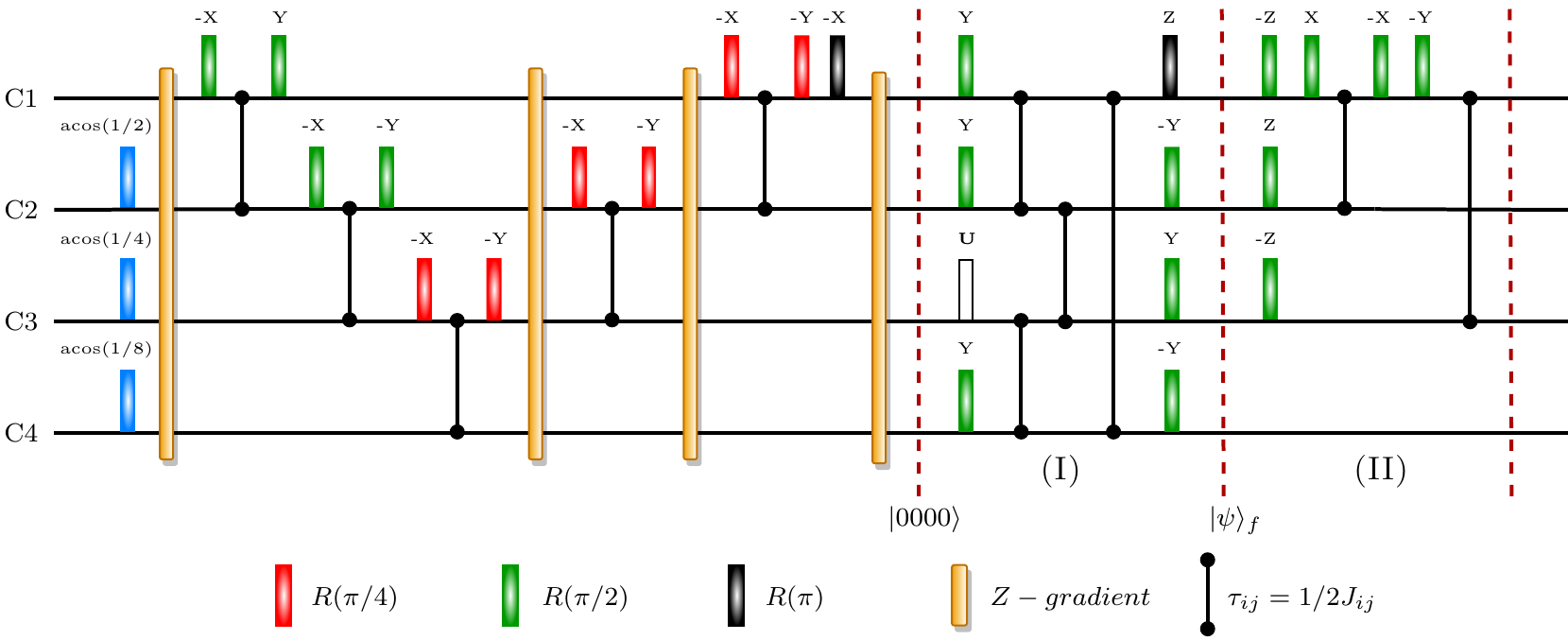}}
\caption{NMR pulse sequence to implement the two-step quantum random walk
circuit to achieve teleportation of an arbitrary single-qubit state
$\vert\phi\rangle$. The sequence of rf pulses, $z$-gradients and time evolution
periods upto the first dashed line prepares the system in the $\rho_{0000}$ PPS,
starting from thermal equilibrium. 
The sequence given in the box labeled 
$\textsc{I}$ implements the two-step quantum walk, 
while
the box labeled $\textsc{II}$ implements the 
unitaries corresponding to the controlled operations.
The phase of each rf pulse is given above each pulse bar;
$\tau_{ij}=\frac{1}{2J_{ij}}$
corresponds to the free evolution period and 
the unitary $U$ is used to
prepare the state $\vert\phi\rangle$; the refocusing schemes
are not depicted.}
\label{fig4}
\end{figure*}

The system was initialized in a pseudopure state (PPS) 
using the spatial averaging method, 
with the density operator being given 
by~\cite{amandeep-qinp,bhole-2020}:
\begin{equation}
 \rho_{0000}=\frac{1-\alpha}{2^4}\mathbb{I}_{16}+\alpha\vert 0000\rangle\langle 0000\vert\label{eq7}
\end{equation}
where $\mathbb{I}_{16}$ is the $16 \times 16$ identity matrix and $\alpha\sim
10^{-5}$ is the thermal spin polarization.  The NMR pulse sequence using rf
pulses, $z$-gradient pulses and time evolution periods used to prepare the
$\rho_{0000}$ PPS starting from thermal equilibrium is depicted in
Figure~\ref{fig4} (upto the first dashed line).  All the rf pulses were
constructed using the GRAPE optimization
technique~\cite{khaneja-jmr-2005,ryan-pra-2008}, which assembles a single shaped
pulse from a large number of individually controlled short segments. All
single-qubit GRAPE pulses were designed to be robust against rf inhomogeneity
with a duration of around 340 $\mu$s - 360 $\mu$s and an average fidelity $\geq
0.997$.  Wherever possible, two or more spin-selective rf pulses were combined
and implemented via a single GRAPE pulse.

Constrained convex optimization based reduced state
tomography~\cite{gaikwad-qip-2021,akshay-scirep}  was used to reconstruct valid experimental
density matrices with a set of 
tomography operations~\cite{li-pra-2017,harpreet-mle,harpreet-pra-2018}:
\begin{eqnarray*} 
&&\{IIII,IIIX,XIIX,XIIY,YYII,IXYY,IYXY,\\ &&IYYY
,XXXI,XYYI,YIXX,YYYI,XXXX,\\ && YXXX,YXYY\} 
\end{eqnarray*}
where $I$ is the identity ('do nothing') operator and $X(Y)$ are qubit-selective
$\pi/2$ rotations applied along the $x(y)$ axis. 

The fidelity of the experimentally reconstructed state as compared to the
theoretically expected state was computed using the
Uhlmann-Jozsa fidelity measure~\cite{uhlmann-rpmp-1976,jozsa-jmo-1994}: 
\begin{equation} 
{\mathcal
F}(\rho_{\text{expt}},\rho_{\text{th}})=\frac{\vert\text{Tr}
[\rho_{\text{expt}}\rho_{\text{th}}^\dagger]\vert}
{\sqrt{\text{Tr}[\rho_{\text{expt}}\rho_{expt}^\dagger]
\text{Tr}[\rho_{\text{th}}\rho_{\text{th}}^\dagger]}}
\label{eq8}
\end{equation}
where $\rho_{\text{expt}}$ and $\rho_{\text{th}}$ are the experimentally
reconstructed and the theoretically expected density matrices, respectively.
The $\vert 0000 \rangle$ PPS was prepared with a total pulse sequence duration
of $\approx 41$~ms with an experimental fidelity $0.9812\pm 0.0020$.

\begin{figure*}[t]
\centering
{\includegraphics[scale=0.9]{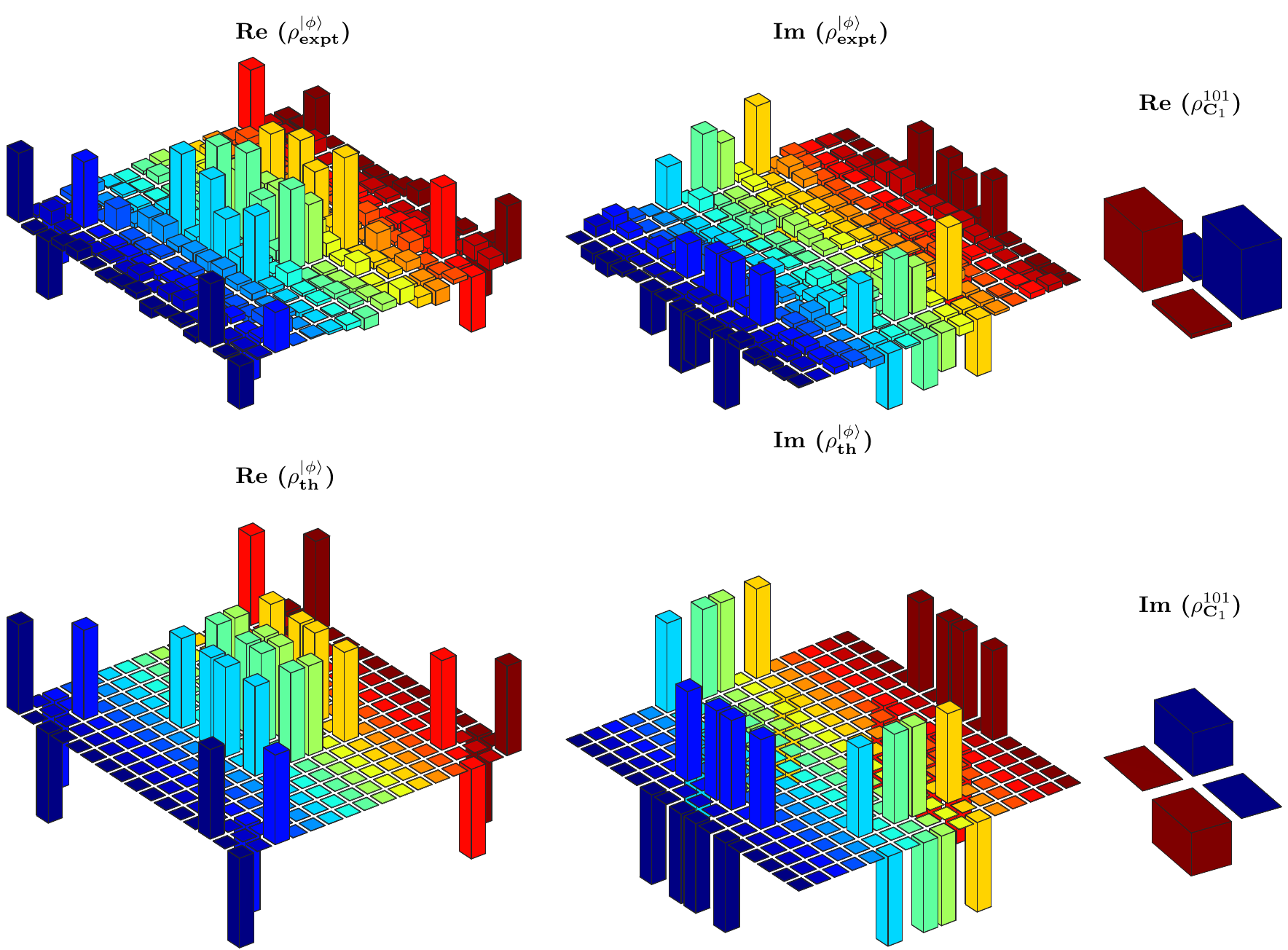}}
\caption{Real (left) and imaginary (right) parts of 
the experimentally tomographed
and the theoretically expected four-qubit state
$\rho_{\text{expt}}^{\vert\phi\rangle}=\vert\psi\rangle_f\langle\psi\vert$ for
the input state $\vert\phi\rangle=\vert -\rangle$, with a 
fidelity of $0.9183 \pm 0.0028$. The tomographs in the right-most column
show the real and
imaginary parts of the teleported state reconstructed at Bob's qubit
$\rho_{\text{C}_1}^{101}$, with a fidelity of $0.9785 \pm 0.002$.
}
\label{fig5}
\end{figure*}
\subsection{Experimental Quantum State Transfer Using Revised Measurements}
\label{sec3.2}
In order to implement the quantum transfer circuit,
we denote the four carbons in the molecule according to the notation: C$_1$ is
Bob's coin, C$_3$ is Alice's coin and C$_2$, C$_4$ are the Arena 
qubits (see Figure~\ref{fig3} for qubit labels). 

The state $\vert\psi\rangle_f$ (Eq \ref{eq4})
after two steps of the random walk is given by:
\begin{eqnarray}
\vert\psi\rangle_f&=&\frac{1}{2} \{(a \vert 0 \rangle+b \vert
1\rangle)\otimes\vert 101 \rangle+(a \vert 0 \rangle-b \vert 1\rangle)
\otimes\vert 111 \rangle+ \nonumber \\
&& (a \vert 1 \rangle+b \vert 0\rangle)\otimes\vert 000 \rangle+ (a \vert 1
\rangle-b \vert 0\rangle)\otimes\vert 010 \rangle\} \label{eq9}
\end{eqnarray}
and the state $\vert\Psi\rangle$ (Eq \ref{eq5}) is given by: 
\begin{equation}
\vert\Psi\rangle=\frac{1}{2} \{\vert 000 \rangle+ \vert 010 \rangle
+ \vert 101 \rangle+ \vert 111 \rangle\}\otimes(a \vert 0 \rangle
+b \vert 1\rangle)
\end{equation}\label{eq10}

We decomposed the quantum circuit given in Figure~\ref{fig2} as a set of
single-qubit rotations and $J$-coupling gates, using controlled-NOT gates (not
shown in the Figure). The corresponding NMR pulse sequence is shown in the
dashed regions $\textsc{I} $ and $\textsc{II} $ in Figure~\ref{fig4}.  This
particular sequence is efficient because of the
cancellation of certain terms in the NMR pulse decomposition
of these unitaries due to spin angular momentum
commutation rules.
We designed gate implementation under simultaneous
evolution of all the $J$-couplings,  using the refocusing method described in
Reference~\cite{bhole-2020}, which finds an optimal solution for the required
interactions using a MATLAB program with the in-built $linprog$ function. In our
case, we found an optimal solution with 9 time periods and 10 refocusing pulses
for simultaneous evolution under the required scalar $J$-couplings $J_{12},
J_{23}, J_{34}$ and $J_{14}$, with a total implementation time of $ 82.19$~ms,
which is 18.2\% shorter  than the 
unoptimized implementation time of
$97.12$~ms.

Furthermore, using this refocusing method and GRAPE optimization, the evolution
time of the coupling $J_{13}$ is dramatically reduced from $\sim 490$~ms to
$\sim 25$~ms.  Our numerically optimized pulse sequence shortens the duration of
the overall pulse sequence as well as reduces the number of pulses used for
refocusing, thus mitigating the undesirable effects of decoherence.  All the
pulses including the refocusing pulses were designed using the GRAPE
optimization algorithm with a fidelity $\geq 0.998$ and are robust against rf
inhomogeneity~\cite{shruti-ijqi}. All the pulses corresponding to single-qubit
rotations have a duration of $\approx 340-360 \mu$s.  

The complete NMR pulse sequence shown in Figure
\ref{fig4} was implemented, with a total time duration of $\sim
156$ ms.  We note here that the measurement outcomes of the
Arena qubits  are either $00$ or $11$ (Table~\ref{table1}).  In case the
Arena qubits are in the $\vert 11\rangle$ state, no further operation is
required.  However, if they are in the $\vert 00 \rangle$ state, the
corresponding controlled 
unitary operation becomes a NOT gate on Bob's qubit, which
is equivalent to implementing a Toffoli gate with zero control. Since both
qubits are in the $\vert 0 \rangle$ state, we considered the control only from
one qubit.  Hence, to simplify the experimental scheme,  instead of implementing
the zero-controlled Toffoli gate, we implemented an equivalent
`zero-controlled-CNOT' gate.  

After implementing the entire pulse sequence, only Bob's qubit (C$_1$) is
tomographed in order to reconstruct the 
transferred state.  This experimental
scheme is efficient in terms of the number of experiments, as it requires only
two tomography pulses $IIII$ and $XIII$ to reconstruct the state.  In order to
evaluate the efficiency of the scheme, we performed the experiment for a set of
four different input states $\vert\phi\rangle=\{\vert 0\rangle,\vert
1\rangle,\vert +\rangle,\vert -\rangle\}$.

\begin{figure}[h] 
\centering 
\includegraphics[scale=1]{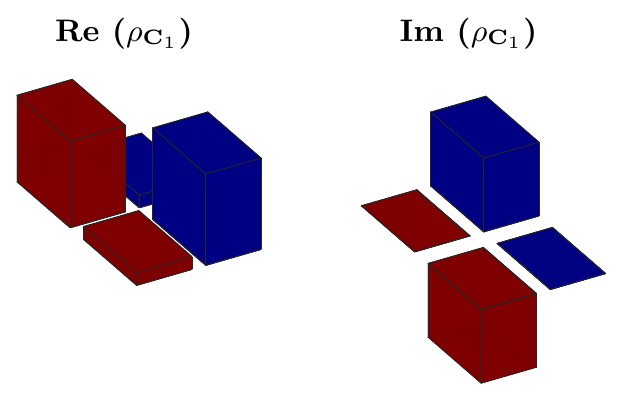}
\caption{Real (left) and imaginary (right) parts of the experimentally
tomographed transferred state at Bob's qubit $\rho_{\text{C}_1}$ with a fidelity
of $0.9896 \pm 0.0007$ for the input state $\vert \phi\rangle=\vert -\rangle$. }
\label{fig7} 
\end{figure}

The experimentally reconstructed transferred state for the input state $\vert
\phi\rangle=\vert -\rangle$ is shown in Figure~\ref{fig7}. The average
experimental fidelities for the input states $\vert 0\rangle,\vert
1\rangle,\vert +\rangle$ and $\vert -\rangle$ are $0.9918\pm 0.0006$, $0.9924\pm
0.0012$, $0.96\pm 0.0027$ and $0.9896\pm 0.0007$, respectively.  These high
state fidelities are a testimony to the fact that the experimental scheme 
is well able to transfer 
the desired single-qubit state.

\subsection{Verifying the Fidelity of the Transferred State}
\label{sec3.3}
We also wanted to independently verify the efficacy of the transfer protocol
by reconstructing the state which has been transferred to Bob and
computing its fidelity. 
We hence repeated the experimental protocol for a set of 
different input states
$\vert\phi\rangle=\{\vert 0\rangle,\vert 1\rangle,\vert +\rangle,\vert
-\rangle\}$, where $\vert +\rangle=\frac{\vert 0\rangle+\vert 1\rangle}{2}$ and
$\vert -\rangle=\frac{\vert 0\rangle+\iota\vert 1\rangle}{2}$.  
The NMR pulse
sequence after PPS preparation and including the box $\textsc{I}$
(Figure~\ref{fig4}) is applied  with a total gate implementation time of $\sim
124$~ms. The density matrix of the four-qubit entangled state
$\rho_{\text{expt}}$ is reconstructed using quantum state tomography. The real
and imaginary parts of the theoretically expected and experimentally
reconstructed four-qubit density matrix for the input state $\vert
\phi\rangle=\vert -\rangle$ are shown in Figure~\ref{fig5}. 

We analyzed this
experimentally reconstructed four-qubit density matrix to obtain  the 
transferred
state.  It is to be noted that the measurement outcomes of Alice's qubits do not
contain any information about the input state $\vert\phi\rangle$.  However, one
can retrieve the input state at Bob's qubit from the results of Alice's
measurements.  To achieve this, we calculated the projections of
$\rho_{\text{expt}}$ onto the basis states $\vert$ C$_2$C$_3$C$_4\rangle=\vert
ijk\rangle=\{\vert000\rangle,\vert010\rangle,
\vert101\rangle,\vert111\rangle\}$~\citep{bau-prl-2012}.  The qubits C$_2$,
C$_3$ and C$_4$ are then traced over and the state of Bob's qubit
$\rho_{\text{C}_1}^{ijk}$ is reconstructed by renormalizing the reduced density
matrix and applying suitable controlled operations $M$ listed in
Table~\ref{table1}: 
\begin{eqnarray}
&\sigma_{\text{C}_1}^{ijk}=
\frac{\text{Tr}_{\text{C}_2\text{C}_3\text{C}_4}[P_{ijk}\rho_{\text{expt}}P_{ijk}^\dagger]}{\text{Tr}[P_{ijk}\rho_{\text{expt}}]}\nonumber\\
&\rho_{\text{C}_1}^{ijk}=M.\sigma_{\text{C}_1}^{ijk}.M^\dagger \label{eq9}
\end{eqnarray} 
where $P_{ijk}=I\otimes\vert ijk\rangle\langle ijk\vert$ are projectors.
Figure~\ref{fig5} depicts the real and imaginary parts of the experimentally
reconstructed state $\rho_{C_1}^{101}$ and the theoretically expected state
respectively, for which the controlled 
operation $M$ is identity.  The fidelities
of the four-qubit state $\rho_{\text{expt}}$ and the reconstructed state of
Bob's qubit corresponding to the different input states $\vert\phi\rangle=\vert
0\rangle,\vert 1\rangle,\vert +\rangle$ and $\vert -\rangle$, are given in
Table~\ref{table2}. 

\begin{widetext}
\begin{center}
\begin{table}[H]
\setlength{\tabcolsep}{12pt} 
\renewcommand{\arraystretch}{1.2}
\footnotesize{
\begin{tabular}{c c c c c c}
\hline \hline
Input State&Fidelity of Four-Qubit&\multicolumn{4}{c}{Fidelity of Bob's 
Reconstructed State }  \\
$\vert\phi\rangle$& State ($\rho_{\text{expt}}$)& $\rho_{\text{C}_1}^{101}$&$\rho_{\text{C}_1}^{111}$ & $\rho_{\text{C}_1}^{000}$& $\rho_{\text{C}_1}^{010}$\\
\hline \hline 
$\vert0\rangle$&$0.9162\pm 0.0025$&$0.9632\pm 0.0022$&$0.9863\pm 0.0008$&$0.9906\pm 0.006$&$0.9651\pm0.0036$\\
$\vert1\rangle$&$0.9180\pm 0.0068$&$0.9809\pm0.0049$&$0.9550\pm00057$&$0.9943\pm 0.0013$&$0.9384\pm 0.0013$\\
$\vert+\rangle$&$0.9357\pm 0.0018$&$0.9880\pm 0.0014$&$0.9928\pm 0.0011$&$0.9891\pm 0.0025$&$0.9775\pm 0.0015$\\
$\vert-\rangle$&$0.9183\pm 0.0028$&$0.9785\pm 0.002$&$0.9527\pm 0.0082$&$0.9712\pm 0.0081$&$0.9666\pm 0.003$
\\
\hline \end{tabular} }
\caption{Fidelities of the experimentally prepared four-qubit state
$\rho_{\text{expt}}$ and the reconstructed state of 
Bob's qubit $\rho_{C_1}^{ijk}$
($ijk = 101,111,000$ and $010$),
for different input states $\vert \phi\rangle$.
}
\label{table2}
\end{table}
\end{center}
\end{widetext}

It is evident that the four-qubit entangled state $\vert\psi\rangle_f$ is a
cluster state~\cite{raussendorf-prl-2001,muralidharan-pra-2008} under local
unitary operations for input states $\vert \pm\rangle$, while it remains
biseparable for input states $\vert 0\rangle$ and $\vert 1\rangle$. We use a
witness operator $\mathcal{W}_{\vert \psi\rangle}=\frac{1}{2} I- \vert
\psi\rangle\langle \psi\vert $ \cite{tokunaga-pra-2006}, to certify the presence
of genuine quadripartite entanglement in $\rho_{\text{expt}}$.  The factor of
$\frac{1}{2}$ in the expression for the witness operator $\mathcal{W}_{\vert
\psi \rangle}$ is the maximum possible fidelity between any biseparable state
and $\vert \psi\rangle$. This ensures that if Tr$[\mathcal{W_{\vert
\psi\rangle}}\rho]<0$, the given state $\rho$ has genuine four-qubit
entanglement, while it is biseparable if Tr$[\mathcal{W_{\vert
\psi\rangle}}\rho]\geq 0$. For the input state $\vert+\rangle$,
Tr$[\mathcal{W_{\vert \psi \rangle_{{\rm f}}}\rho_{{\rm expt}}}]
=-0.4358\pm0.0018$ 
and for
the input state $\vert-\rangle$, Tr$[\mathcal{W_{\vert
\psi \rangle_{{\rm f}}}\rho_{{\rm expt}}}]=-0.4183\pm 0.0028$, 
which clearly indicates the
presence of genuine quadripartite entanglement.

\begin{figure}[H]
\centering
{\includegraphics[scale=0.85]{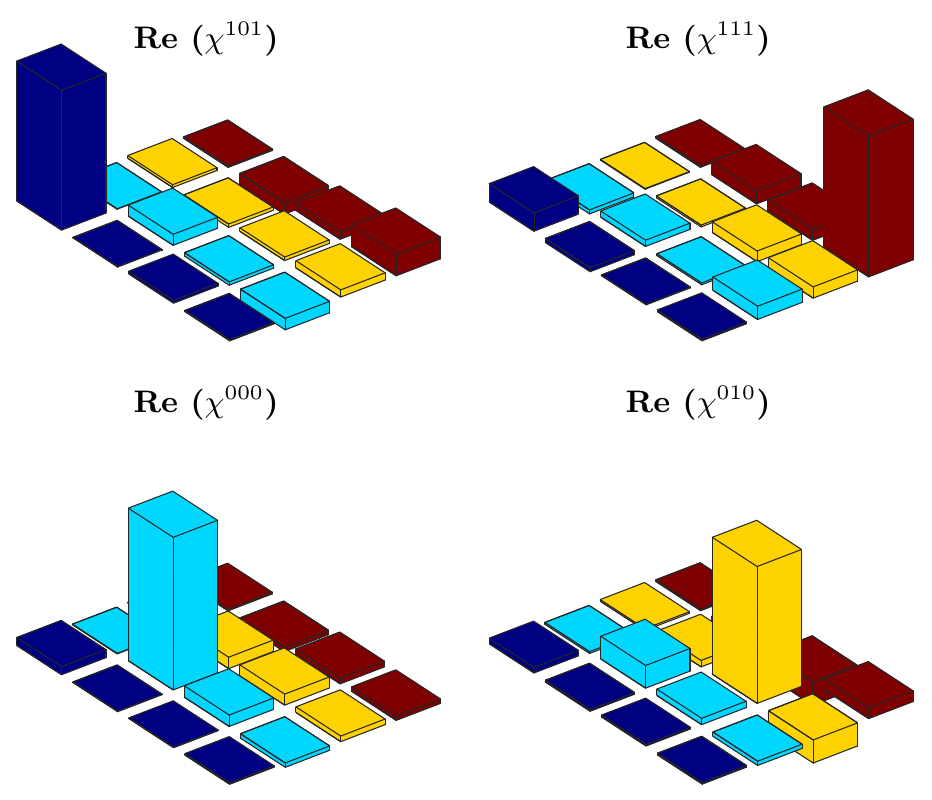}}
\caption{Real parts of the experimentally tomographed process matrix $\chi$ for
different projections of $\rho_{\text{expt}}$ onto the basis states
$\vert000\rangle,\vert010\rangle,\vert101\rangle$ and $\vert111\rangle$.}
\label{fig6}
\end{figure}

We performed constrained convex optimization based quantum
process tomography~\citep{gaikwad-qip-2021} to fully characterize the
state transfer protocol. 
One can represent a completely positive map 
$\rho_\phi=\vert\phi\rangle\langle\phi\vert$ 
with a
quantum state 
$\phi$, which is to be transferred.
In the operator sum representation, the transfer 
process
is defined as~\cite{gaikwad-pra-2018}:
\begin{equation}
\mathcal{E}(\rho_\phi)=\rho_{C_1}^{ijk}=\sum_{m,n=1}^{d^2} \chi_{mn}^{ijk} E_m
\rho_\phi E_n^\dagger \label{eq10}
\end{equation}
where $d$ is the dimension of 
the Hilbert space, $E_m$ refers to the Pauli operator basis
$\{I,\sigma_x,\sigma_y,\sigma_z\}$ and $\chi_{mn}^{ijk}=\sum_\alpha a_{\alpha m}
a_{\alpha n}^*$ is the process matrix, which is 
a positive Hermitian matrix and
satisfies the trace-preserving constraint $\sum_{mn}\chi_{mn}^{ijk} E_n^\dagger
E_m=I$. Figure~\ref{fig6} shows the tomographed process matrices $\chi^{ijk}$
which verify the expected unitary 
controlled operations listed in Table~\ref{table1}. The
fidelity of 
the experimentally reconstructed process matrix $\chi^{ijk}_{\text{expt}}$
with respect to the theoretically expected process
matrix  $\chi^{ijk}_{\text{th}}$ is calculated using
the measure~\cite{gaikwad-qip-2022}:
\begin{equation}
{\mathcal
F}^{ijk}=\frac{\vert\text{Tr}[\chi^{ijk}_{\text{expt}}\chi^{ijk^\dagger}_{\text{th}}]\vert}{\sqrt{\text{Tr}[\chi^{ijk}_{\text{expt}}\chi^{ijk^\dagger}_{\text{expt}}]\text{Tr}[\chi^{ijk}_{\text{th}}\chi^{ijk^\dagger}_{\text{th}}]}}\label{eq11}
\end{equation}

The corresponding process fidelities for different projections are $\mathcal
{F}^{101}=	0.9682\pm 0.0021, \  \mathcal {F}^{111}=	0.9658\pm
0.0015, \ \mathcal {F}^{000}=	0.9842\pm 0.0023, \   $ and $\mathcal {F}^{010}=
0.9450\pm 0.0015$.

Our results lead to the conclusion that
the experimental scheme is able to
achieve almost perfect transfer of 
an arbitrary single-qubit state with high fidelity. 
We note
here that for the NMR implementation of the state
transfer scheme given in Figure~\ref{fig2}, we have replaced single-shot
readout~\cite{mallet} and real-time feedback~\cite{doherty} required for
state transfer, with controlled unitary operations
based on Alice's measurement outcomes.  These controlled unitaries posed
a significant experimental challenge 
since some J couplings between qubits in the
molecule chosen for the experiments are of a very weak strength. This implies
that the corresponding gate times are rather long, increasing the likelihood of
decoherence adversely affecting the computation, thereby leading to lower
fidelities of the transferred state.  For instance, the total time to naively
implement the teleportation circuit for weakly coupled qubits C1 and C3 (J$_{13}
\approx 1.02$ Hz) was around $490 \mu$s; using specially crafted and
GRAPE-optimized pulses we were able to dramatically reduce the experimental
implementation time to around $25 \mu$s. This allowed us to experimentally
implement the state transfer protocol 
with high fidelity.  
\section{Conclusions}
\label{sec4}
We experimentally demonstrated a quantum state
transfer
scheme between two parties, Alice and Bob, using a two-step
quantum walk on a cycle with four vertices. 
We
performed constrained convex optimization based quantum
state tomography and reconstructed the experimental density
matrix.  The creation of genuine
quadripartite entanglement during the quantum walk was
verified by using an entanglement witness. The
transferred
state was reconstructed on Bob's qubit by calculating its
projection onto the basis states of Alice's qubit. Quantum
process tomography was also performed to characterize the
process matrix of the state transfer
protocol, 
and only Bob's qubit was measured in order to reconstruct
the transferred state. 
The scheme was able to achieve near
perfect transfer of an arbitrary single-qubit state,
with high state fidelity. Furthermore, the experimental
circuit alongwith a few local unitaries can be used to
generate four-qubit cluster states, which are of much
interest in quantum information processing.

Our experimental schemes are general and can be
easily extended to transferring states 
of a larger qubit
register size via multi-coin, multi-walker setups.  Our
results demonstrate that coined quantum random walk schemes
can be used to achieve robust transfer of an
arbitrary quantum state and pave the way for wider applications in
quantum communication and quantum information processing.

\begin{acknowledgments}
All experiments were performed on a Bruker Avance-III 600
MHz FT-NMR spectrometer at the NMR Research Facility at
IISER Mohali.  Arvind acknowledges funding from the
Department of Science and Technology (DST), India, Grant
No:DST/ICPS/QuST/Theme-1/2019/Q-68.  K.D.  acknowledges
funding from the Department of Science and Technology (DST),
India, Grant No:DST/ICPS/QuST/Theme-2/2019/Q-74.  G.S.
acknowledges University Grants Commission (UGC), India, for
financial support.
\end{acknowledgments}


%
\end{document}